\documentclass[aps,prd,twocolumn,superscriptaddress]{revtex4-2}
\usepackage{amsmath}
\usepackage{amssymb} 
\usepackage{amsthm}
\usepackage{graphicx}
\usepackage{dcolumn}
\usepackage{bm}
\usepackage{braket}
\usepackage{xcolor}
\usepackage[normalem]{ulem}
\usepackage[colorlinks=true,urlcolor=blue,citecolor=blue,linkcolor=blue]{hyperref}
\usepackage[capitalise]{cleveref}
\usepackage{physics}
\usepackage{tabularray}
\usepackage{bbm}
\usepackage{upgreek}
\usepackage{tabularray}

\begin{document}

\title{Large-scale implementation of quantum subspace expansion with classical shadows}

\date{\today}

\author{Laurin E. Fischer}
\email{aur@zurich.ibm.com}
\affiliation{IBM Quantum, IBM Research Europe -- Zurich, 8803 R\"{u}schlikon, Switzerland}
\affiliation{Theory and Simulation of Materials, {\'E}cole Polytechnique F{\'e}d{\'e}rale de Lausanne, 1015 Lausanne, Switzerland}

\author{Daniel Bultrini}
\affiliation{Theoretische Chemie, Physikalisch-Chemisches Institut, Heidelberg University, Heidelberg, Germany.}

\author{Ivano Tavernelli}
\affiliation{IBM Quantum, IBM Research Europe -- Zurich, 8803 R\"{u}schlikon, Switzerland}

\author{Francesco Tacchino}
\email{fta@zurich.ibm.com}
\affiliation{IBM Quantum, IBM Research Europe -- Zurich, 8803 R\"{u}schlikon, Switzerland}

\date{\today}

\begin{abstract}
Quantum subspace expansion (QSE) offers promising avenues to perform spectral calculations on quantum processors but comes with a large measurement overhead. 
Informationally complete (IC) measurements, such as classical shadows, were recently proposed to overcome this bottleneck.
Here, we report the first large-scale implementation of QSE with IC measurements.
In particular, we probe the quantum phase transition of a spin model with three-body interactions, for which we observe accurate ground state energy recovery and mitigation of local order parameters across system sizes of up to 80 qubits. 
We achieve this by reformulating QSE as a constrained optimization problem, obtaining rigorous statistical error estimates and avoiding numerical ill-conditioning.
With over $3\times10^4$ measurement basis randomizations per circuit and the evaluation of $\mathcal{O}(10^{14})$ Pauli traces, this represents one of the most significant experimental realizations of classical shadows to date.
\end{abstract}

\maketitle

\textit{Introduction---}Spectral calculations are a key target for many-body quantum simulation frameworks. 
Until recently, heuristic approaches based on the \emph{variational quantum eigensolver} (VQE) dominated the near-term quantum computing literature on the topic, thanks to their relatively low demands on quantum processor performance~\cite{tilly2022variational}. 
However, practical issues connected to trainability~\cite{larocca2025barren} and large measurement overheads~\cite{gonthier2020identifying} have so far hindered their scalability beyond small proof-of-principle demonstrations. 
In response to this, alternative methods for calculating ground and excited states are being explored.
A promising class of post-VQE quantum algorithms for spectral calculations is \emph{quantum subspace expansion} (QSE)~\cite{Motta_2024}. 
Assume that, as usual, the goal is to find the energy $E_{\text{gs}}$ of a ground state $\ket{\psi_{\text{gs}}}$ of a target Hamiltonian $H$, and suppose that a certain ``root state'' $\rho_0$, close to $\ket{\psi_{\text{gs}}}$ can be prepared on a quantum computer.
QSE relies on the fact that, while $\rho_0$ alone may not yield the desired accuracy, e.g., due to noise or unitary circuit approximations, it might still represent a useful resource.
In fact, there exist several ways in which $\rho_0$ can be employed to inform a subsequent classical calculation.
These include, for instance, matrix diagonalization based on projection into a space spanned by individual samples~\cite{kanno2023quantum, robledomoreno2025}, the extraction of an effective representation of $\rho_0$ based on classical shadows~\cite{huang2020predicting}, and subsequent processing of the quantum computer outputs via Tensor Networks~\cite{garcia-perezvirtuallinearmap2022, filippov2023scalable} or Neural Quantum States~\cite{NN_postprocessing}.

In QSE, a space of states is spanned on top of $\rho_0$ with $L$ Hermitian expansion operators $\sigma_i$ as~\cite{yoshioka2022generalized}
$\rho_\text{SE} (\vec{c}) = W^\dagger \rho_0 W / \Tr[W^\dagger \rho_0 W]$, with $W = \sum_{i=1}^L c_i \sigma_i$, parametrized by a coefficient vector $\vec{c} = \left(c_1, \dots, c_L\right)$. 
This can be regarded as accessing additional states beyond the scope of the quantum processor~\cite{takeshita2020increasing, urbanek2020chemistry}, and may also serve as noise-agnostic error mitigation~\cite{bonetmonroig2018lowcosta, cai2021quantum,suchsland2021algorithmic,yoshioka2022generalized}. 
The expectation value of an observable $O$ over a state in the subspace then becomes 
\begin{align}
\label{eqn:exp_value_subspace}
\Tr[O \rho_\text{SE}(\vec{c}) ] = \frac{\sum_{i,j=1}^L c_i^* c_j \mathcal{O}_{ij}}{\sum_{i,j=1}^L c_i^* c_j\mathcal{S}_{ij} }
,
\end{align}
where $\mathcal{O}_{ij} = \Tr[\sigma_i^\dagger \rho_0 \sigma_j O ] \quad \text{and} \quad \mathcal{S}_{ij} = \Tr[\sigma_i^\dagger \rho_0 \sigma_j]$.
For ground state search, the goal is to find the minimal expectation value $\operatorname*{min}_{\vec{c}} \Tr[H \rho_\text{GSE}(\vec{c})]$ which corresponds to the smallest pseudoeigenvalue $\lambda$ of the generalized eigenvalue problem 
$\mathcal{H} \vec{c} = \lambda \mathcal{S} \vec{c}$ where $\mathcal{H}$ is the subspace-projected Hamiltonian and $\mathcal{S}$ is called the overlap matrix.
For reasonable values of $L$, this is tractable on a classical computer by inverting the overlap matrix and computing the smallest eigenvalue of $\mathcal{S}^{-1} \mathcal{H}$.
For ground state calculations, the powers of the Hamiltonian $H^p$ are particularly suitable expansion operators. 
Indeed, the overlap of the states $H^p \rho_0 {H^p}^\dagger$ (once properly normalized) with the true ground state increases exponentially with $p$, provided $\rho_0$ has non-zero overlap with $\ket{\psi_{\text{gs}}}$. 
This corresponds to the well-known Krylov subspace diagonalization method~\cite{motta2020determining,suchsland2021algorithmic,lanczos_real_time,kirby2023exact}. 

The main experimental cost of QSE stems from the number of measurements required to estimate all matrix entries $\mathcal{O}_{ij}$ and $\mathcal{S}_{ij}$, which depends on the dimension $L$ and the complexity of the expansion operators.
For approaches that measure each observable individually (or in commuting groups) this can be a significant roadblock for scaling in practice~\cite{ollitrault2020quantum,selisko2024dynamical}. 
To alleviate this, \emph{informationally complete} (IC) measurements, such as \emph{classical shadows}~\cite{huang2020predicting} (CS) can be used to estimate all underlying observables from one set of measurement samples~\cite{garcia-perez_learning_2021, choi2023measurementa}.
CS measurements have recently been realized in a series of large-scale experiments~\cite{fischer2024dynamical, votto2025learning} and the use of CS to systematically build high-dimensional subspaces spanned by low-weight Paulis has been proposed in Ref.~\cite{boyd2025high}.
However, while the latter offered a promising avenue, its statistical analysis focused on worst-case upper bounds rather than reliable statistical error bars for the resulting estimators. 
Furthermore, no experimental demonstration of such ideas has been achieved so far.
In this letter, we close both gaps by formulating subspace expansion as a constrained optimization problem.
In contrast to previous methods, this allows us to obtain accurate statistical uncertainties and provides an external tuning knob to control these errors, trading potential bias for variance. 
Leveraging this technique, we experimentally demonstrate noise-agnostic error mitigation via QSE in ground state preparation circuits of spin models for systems with up to 80 qubits.

\textit{Theory---}
An IC measurement is described by a set of $n$ positive semi-definite Hermitian operators $\{M_k\}_{k \in \{1, \dots, n \}}$ with $\sum_{k=1}^n M_k = \mathbbm{1}$ such that any observable $O$ can be expressed as $O = \sum_{k=1}^{n} \omega_k M_k$ for some $\omega_k \in \mathbb{R}$~\cite{fischer2022ancillafree}. 
The expectation value $\langle O \rangle$ then becomes $\Tr[\rho_0 O] = \sum_k \omega_k \Tr[\rho_0 M_k] = \mathbb{E}_{k \sim p_k}[\omega_k]$, i.e., $\expval{O}$ can be expressed as the mean value of the random variable $\omega_k$ over the probability distribution $\{p_k  = \Tr[\rho_0 M_k]\}$.
Given a sample of $S$ measurement outcomes $\{ k^{(1)}, \dots, k^{(S)} \}$, we can thus construct an unbiased estimator of $\expval{O}$ as
\begin{equation}
\label{eqn:canonical_estimator}
    \hat{o} : \{k^{(1)},\dots, k^{(S)}\} \mapsto \frac{1}{S} \sum_{s=1}^{S} \omega_{k^{(s)}}.
\end{equation}
The statistical variance of this estimator is given by the standard error on the mean $\mathrm{Var}[\hat{o}] = \mathrm{Var}[\omega_k] / S$.
The coefficients $\omega_k$ can be obtained from frame theory~\cite{innocenti2023shadow} where each POVM operator $M_k$ is associated with a dual operator $D_k$ (also known as the ``classical shadow'' of a state) such that $\omega_k = \Tr[O D_k]$. 

Given $S$ POVM outcome samples $\{k^{(1)}, \dots , k^{(S)}\}$ measured from the base state $\rho_0$, we aim to simultaneously estimate all matrix entries of the projected Hamiltonian $\mathcal{H}_{ij}$ and the overlap matrix $\mathcal{S}_{ij}$. 
Following Eq.~\eqref{eqn:canonical_estimator}, a canonical estimator for each matrix element $\mathcal{H}_{ij}$ and $\mathcal{S}_{ij}$ is obtained by simply replacing $\rho_0$ in each expression with the corresponding dual operator $D_k$ and then averaging over all shots. 
In principle we could estimate all entries of $\mathcal{H}$ and $\mathcal{S}$ this way and then solve the generalized eigenvalue problem. 
However, in practice this direct inversion of the overlap matrix can be numerically unstable due to ill-conditioning of the matrix under shot noise and experimental imperfections, particularly when the expansion operators are non-orthogonal~\cite{epperly2021theory}.
This issue is often addressed through regularization of the overlap matrix, e.g., by discarding dimensions that correspond to small singular values of $\mathcal{S}$~\cite{urbanek2020chemistry}. 
Furthermore, since with IC measurements we do not estimate the matrix entries independently, additional covariances between them need to be accounted for when deriving statistical error bars.

\begin{figure}
    \centering
    \includegraphics[width=\columnwidth]{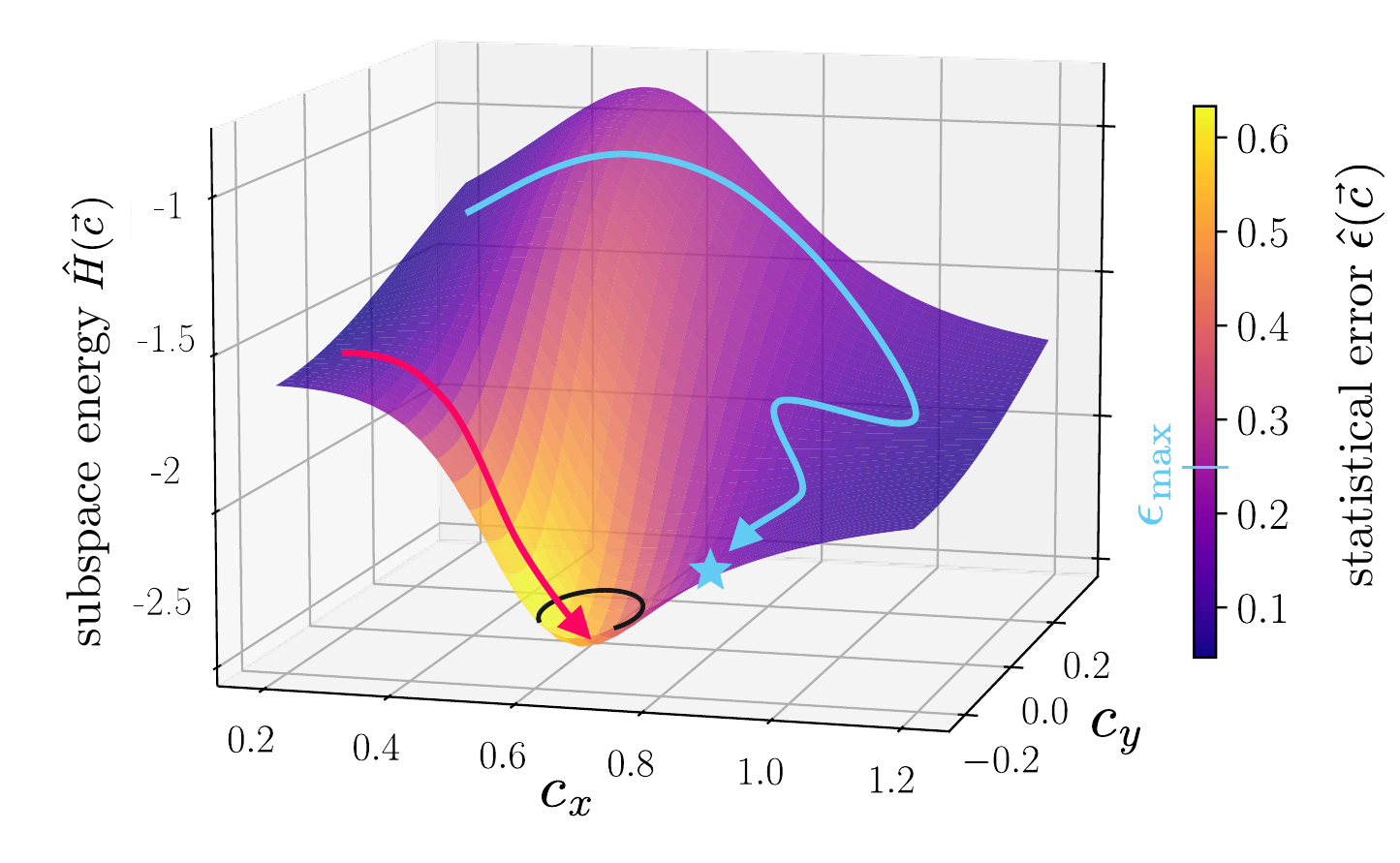}
    \caption{
    {Subspace expansion as a constrained optimization problem.} The z axis shows the estimated energy $\hat{H}(\vec c) $ when varying two dimensions of the subspace coefficients $\vec{c}$, while color indicates the statistical error $\hat{\epsilon}(\vec{c})$ of the energy estimation. Naive minimization of the energy (red arrow) may yield an energy with high statistical error that can severely violate the variational principle (black circle indicates the true ground state energy). Constrained optimization with a maximum allowed error $ \epsilon_\text{max}$ avoids statistically unstable regions (blue path). Data corresponds to experiments presented in Fig.~\ref{fig:PSE_MPS_results} for $N=48$ and $g=-0.5$.}
    \label{fig:energy_landscape}
\end{figure}

Here, we present a different scheme to avoid ill-conditioning which is tailored to handle such inherent covariances.
Assume that we have an estimator $\hat{H}(\vec{c})$ of the energy $\Tr[H\rho_\text{SE}(\vec{c}) ]$ of the state at subspace vector $\vec{c}$ and that we can further estimate the statistical error of $\hat{H}(\vec{c})$ as 
 $\hat{\epsilon}(\vec{c})$.
In practice, we would like to compute the lowest possible subspace energy given a maximally tolerated error $\epsilon_\text{max}$.
This is achieved through the constrained optimization problem
\begin{equation}
\label{eq:PSE_constrained_optimization}
\underset{\vec{c} \in \mathbbm{R}^L}{\arg\min}\ \hat{H}(\vec{c}) \quad \text{subject to} \quad \hat{\epsilon}(\vec{c}) \leq \epsilon_\text{max}.
\end{equation}
This allows us to explore only those regions of the subspace where the statistical error is well controlled. 
In contrast to regularization techniques, it offers a direct tuning knob to trade off statistical errors for potentially lower estimates of the energy, see Fig~\ref{fig:energy_landscape} and Appendix~\ref{app_constrained_optimization} for details.

We now construct the estimators $\hat{H}(\vec{c})$ and $\hat{\epsilon}(\vec{c})$ based on IC samples. 
Let $x_k =\sum_{i,j=1}^L c_i^* c_j \Tr\bigr[D_k \sigma_j H \sigma_i^\dagger \bigr]$ and $y_k = \sum_{i,j=1}^L c_i^* c_j \Tr\bigr[D_k \sigma_j \sigma_i^\dagger \bigr]$ be the random variables of the numerator and denominator of Eq.~\eqref{eqn:exp_value_subspace} distributed according to $p_k$.
Then the subspace expectation value $ \Tr[H \rho_\text{GSE}(\vec{c})] = \mu_x /\mu_y$ is given by the ratio of $\mu_x = \mathbbm{E}_{k\sim p_k}[x_k]$ and $\mu_y = \mathbbm{E}_{k\sim p_k}[y_k]$. 
An asymptotically-unbiased and consistent estimator of this ratio is given by the ratio of the sample means~\cite{van_kempen_mean_2000}
\begin{equation}
\label{eq:ratio_ostimator_O}
\hat{H}(\vec{c}): \{k^{(1)}, \dots , k^{(S)}\} \mapsto \frac{\bar{x} }{\bar{y}} 
\end{equation}
with $\bar{x} = \sum_{i=1}^S x_{k^{(i)}} / S$ (similarly for $\overline{y}$).
While no closed-form expression exists for the variance $\epsilon^2(\vec{c}) = \mathrm{Var}[\hat{H}(\vec{c})]$ of this estimator, it can be well approximated~\cite{huggins2021virtual} by a second-order Taylor approximation around $(\mu_x, \mu_y)$ which yields~\cite{van_kempen_mean_2000}
\begin{align}
\label{eq:ratio_estimator_variance}
\scalebox{0.96}{$\epsilon^2(\vec{c}) \approx \dfrac{1}{S} \left( \dfrac{\mathrm{Var}[x_k]}{\mu_y^2} +\dfrac{\mu_{x}^{2}\mathrm{Var}[y_k]}{\mu_{y}^{4}}-\dfrac{2\mu_{x}\mathrm{Cov}[x_k,y_k]}{\mu_{y}^{3}}\right)$}.
\end{align}
Note that for $\sigma_0 = \mathbbm{1}$ and $\vec{c} = \left(1, 0, 0, \dots\right)$ this reduces to the standard error on the mean.
We obtain the estimated error $\hat{\epsilon}(\vec{c})$ by inserting the sample means $\bar{x}$ for $\mu_x$, the sample standard deviations $\sum_{i=1}^S |x_{k^{(i)}} - \bar{x}|^2 / (S-1)$ for $\mathrm{Var}[x_k]$ (similarly for $y_k$) and the sample covariance $ \sum_{i=1}^{S} (x_{k^{(i)}} - \bar{x})(y_{k^{(i)}} - \bar{y}) / (S-1)$ for $\mathrm{Cov}[x_k,y_k]$ into Eq.~\eqref{eq:ratio_estimator_variance}. 
The covariance term captures the contributions to the statistical error that arise from re-using the same measurement samples to estimate all matrix entries. 

The above statistical error propagation holds when the POVM samples are independent and identically distributed (i.i.d.). 
In practice, for POVMs from randomized measurements (such as CS), it is often experimentally convenient to take multiple measurement shots for each randomized readout basis, which violates the i.i.d. assumption.
In this case, we can redefine the random variables $x_k$ and $y_k$ as their averaged value over all repeated shots taken in the same measurement basis. 
In this way, the obtained values for $x_k$ and $y_k$ remain i.i.d., and Eqs.~\eqref{eq:ratio_ostimator_O}--\eqref{eq:ratio_estimator_variance} remain valid for solving the constrained optimization problem from Eq.~\eqref{eq:PSE_constrained_optimization}. 

Besides substantial efficiency gains from evaluating the matrix entries of the subspace expansion in parallel rather than sequentially~\cite{choi2023measurementa}, our approach based on IC measurements offers two additional advantages over traditional methods. 
Firstly, the expansion operators $\{\sigma_1, \dots, \sigma_L\}$ can be chosen \textit{a posteriori} after measurements have been taken, as also pointed out by Ref.~\cite{boyd2025high}. 
Thus, the same measurements can be reused to iteratively optimize the chosen expansion operators to minimize the energy while the constrained optimization guarantees that the variational principle is respected up to the chosen statistical precision $\epsilon_\text{max}$. 
Secondly, the framework offers the flexibility of estimating additional observables $O$ for the obtained energy-optimized state without additional measurements by constructing the estimator $\hat{O}(\vec{c}_\text{opt})$ equivalently to Eq.~\eqref{eq:ratio_ostimator_O} (replacing all occurrences of $H$ with $O$) where $\vec{c}_\text{opt}$ is the solution to Eq.~\eqref{eq:PSE_constrained_optimization}. 

\begin{figure*}
    \centering
    \includegraphics[width=1\textwidth]{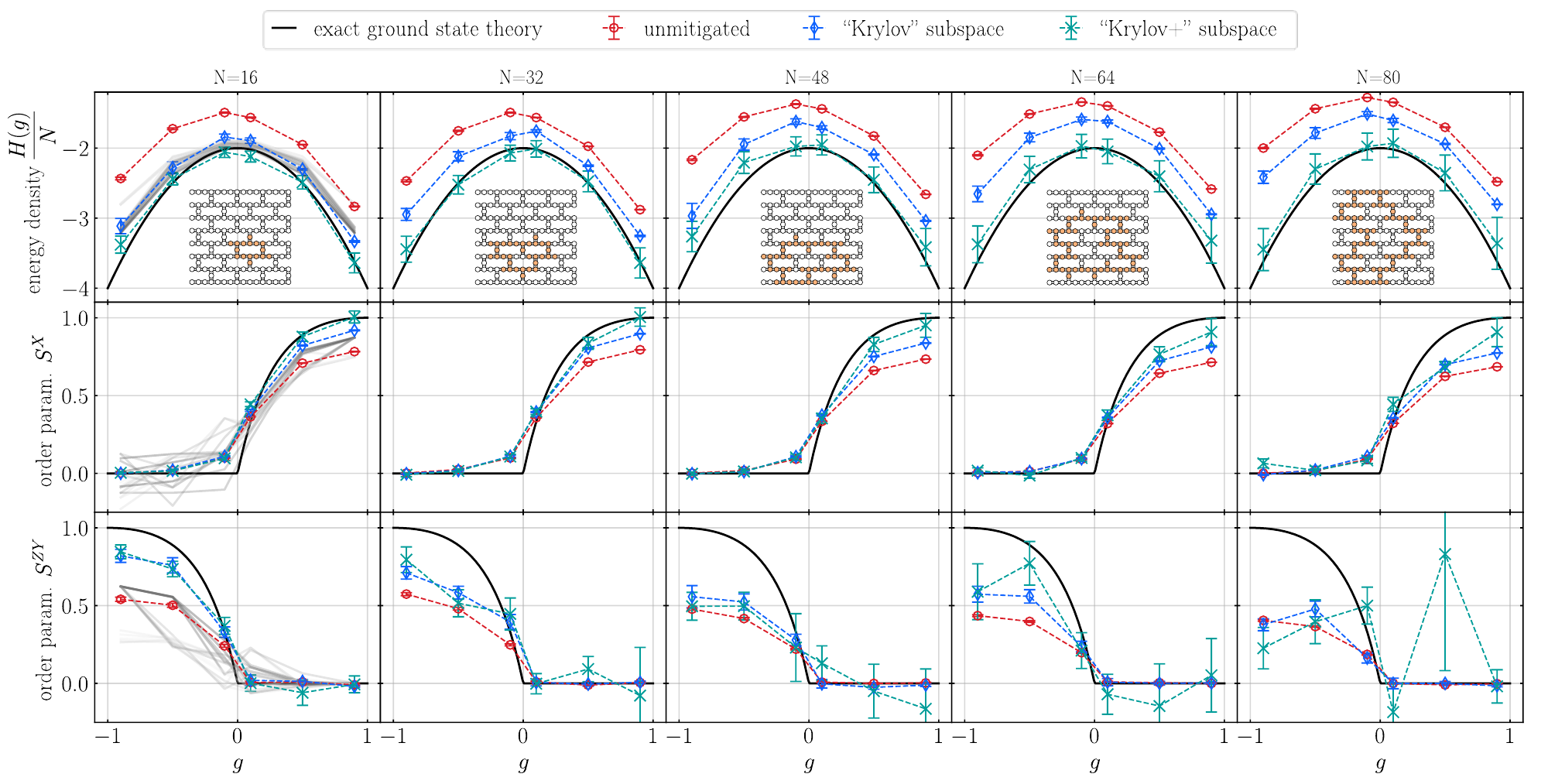}
    \caption[]{Error mitigation with subspace expansion based on IC measurements of ground state preparation circuits for a spin model.
    The qubit number increases from left to right from $N=16$ to $N=80$ while the three rows show the energy density and the order parameters of the traversed phase transition. 
    Error bars represent the estimated statistical error $\hat \epsilon (\vec c)$ and the inlays show the chosen physical qubits of the 156-qubit device \textit{ibm\_fez}. 
    For $N=16$, grey shaded curves show the theoretical values for the lowest 25 excited states obtained with exact diagonalization, with a color gradient from black for the ground state to lighter shades for progressively higher-excited states.
}
\label{fig:PSE_MPS_results}
\end{figure*}

\textit{Experimental demonstration---} We experimentally demonstrate our method by addressing the task of recovering the ground state energy of a Hamiltonian with known approximate ground state preparation circuits. 
We consider a one-dimensional spin chain originally studied in Ref.~\cite{MPS_toy_model_paper} described by the Hamiltonian (with periodic boundary conditions)
\begin{equation}
\label{eq:Hamiltonian_MPS_toy_model}
H = \sum_{i=1}^N \left( - g_{zz} Z_i Z_{i+1} - g_x X_i + g_{zxz} Z_i X_{i+1} Z_{i+2}  \right)
\end{equation}
where the parameter trajectory (for $ g \in \left[ -1, +1 \right]$) 
$\left( g_{zz}(g), g_x(g), g_{zxz}(g) \right) = \left(2(1- g^2), (1+g)^2, (g-1)^2 \right)$ traverses a quantum phase transition with a critical point at $g=0$ between a symmetry-protected topological (SPT) phase where the $ZXZ$-terms dominate, and a trivial phase where the $X$-terms dominate. 
Along this trajectory, the ground state of the model corresponds to a translation-invariant normal matrix product state (MPS)~\cite{MPS_system_first_demo} with an energy density $E_{\text{gs}} / N = -2\left(g^2+1\right)$.
We implement the latter as a quantum circuit leveraging a recently introduced algorithm based on the renormalization-group transformation that approximately prepares the MPS with an error $\delta$ in depth $\mathcal{O}\left(\log(N/\delta)\right)$~\cite{MPS_preparation_theory}. 
The approximation is determined by a ``blocking number'' $q$ which navigates a trade-off between circuit depth and the fidelity of the prepared state. 
Here, we opt for a blocking number of $q=4$ which has been shown to be a reasonable choice for current noisy quantum hardware~\cite{scheer2025renormalization}. 

We implement the MPS ansatz circuits with system sizes increasing from $N=16$ with 108 entangling CZ gates to $N=80$ with 540 CZ gates on the IBM Quantum superconducting qubit device \emph{ibm\_fez} of the Heron R2 generation, 
see Appendix~\ref{app_device} for details. 
We measure an IC POVM that consists of uniformly randomized single-qubit $X$, $Y$, and $Z$ measurements (standard local classical shadows with canonical duals~\cite{innocenti2023shadow}, see also Appendix~\ref{app_classical_shadows}).
For each studied system size $N$ and parameter value $g$, we perform measurements in $32,768$ bases with $8$ shots per basis.
This POVM can efficiently estimate observables of low Pauli weight $w$ with statistical errors increasing exponentially with $w$~\cite{huang2020predicting}.
Hence, in principle, we could now choose any set of sufficiently local observables as expansion operators $\sigma_i$ of the subspace. 
We first select $\sigma_1 = \mathbbm{1}$ to ensure that the original state $\rho_0$ is part of the subspace. In this way, the subspace-optimized energy can never be worse than the measured one. 
Next, we choose $\sigma_2 = H$ to boost the ground state population in the spirit of Krylov subspaces.
Along these lines, one would then ideally like to add more powers of the Hamiltonian as expansion operators. 
However, in our model $\sigma_2 = H$ already leads to a term $\mathcal{H}_{2, 2} = \Tr[\rho_0 H^3]$ which is an observable with $2.14 \times 10^6$ Pauli terms at $N=80$. 
The trace of each of these Pauli terms with each measured dual operator (which are $262,144$ in our case) needs to be evaluated. 
The inclusion of $\sigma_3 = H^2$ would result in a term $\mathcal{H}_{3, 3} = \Tr[\rho_0 H^5]$ with $35\times 10^9$ Pauli terms and strings with weight up to $w=15$. 
These both exceed our available classical processing resources and are beyond the scope of estimation with single-qubit classical shadows. 
We hence limit ourselves to one Krylov dimension and label such expansion sets as $\text{``Krylov''}\,\, = \{\mathbbm{1}, H\}$.

Instead of increasing the Krylov dimension, we further lower the energy by systematically adding low-weight Pauli expansion operators with a strategy proposed similarly in Ref.~\cite{boyd2025high}.
For Pauli weights $w \in \{1, 2, 3, 4, 5\}$, we sample 150 Pauli strings of weight $w$ uniformly at random to form a pool $\mathcal{P}_w$. 
For each $P_i \in \mathcal{P}_w$, we perform a two-dimensional subspace expansion with operators $\{\mathbbm{1}, P_i \}$.
We then rank the Pauli strings within each pool according to the amount by which they reduce the energy relative to the unmitigated estimate, when optimizing for the lowest upper error given by
$ \arg\min_{\vec{c}}\, \hat{H}(\vec{c}) + \hat{\epsilon}(\vec{c})$. 
Finally, we include up to five of the best-performing Paulis for each considered weight into the set of expansion operators, see Appendix~\ref{app_expansion_ops}. 
Together with $\sigma_1 = \mathbbm{1}$ and $\sigma_2 = H$, this results in a subspace which we refer to as ``Krylov+''.
The total numbers of single-qubit Pauli traces of the form $\Tr[P_i P_j], P_i,P_j\in \{I, X, Y, Z\}$ between dual operators of measurement outcomes and the observables of the subspace range from $4.3 \times 10^{11}$ for $N=16$ to $5.6\times 10^{13}$ for $N=80$, see Appendix~\ref{app_classical_shadows}. 
For both considered subspaces, we solve Eq.~\eqref{eq:PSE_constrained_optimization} with the \emph{constrained optimization by linear approximation} (COBYLA) algorithm~\cite{Powell1994}.
The allowed maximum errors $\epsilon_\text{max}$ were chosen as a fixed multiple of the unmitigated signal to ensure a signal-to-noise ratio ranging from $5\,\%$ for $N=16$ to $15\,\%$ for $N=80$, see Appendix~\ref{app_constrained_optimization}.

The results for the error-mitigated energy densities of the MPS ground state circuits for $g\in \{-0.9, -0.5, -0.1, 0.1, 0.5, 0.9\}$ are shown in the top row of Fig.~\ref{fig:PSE_MPS_results}.
Across all studied system sizes, the ``Krylov'' subspace already significantly improves on the unmitigated energies, while the ``Krylov+'' approach manages to accurately recover the true ground state expectation values while respecting the variational principle within statistical uncertainties. 
As the system size increases, the statistical errors of the unmitigated results become progressively worse which leads to larger error bars also for the mitigated values.
For the smallest system size of $N=16$ we can perform an exact diagonalization of the Hamiltonian to get the full low-energy (density) spectrum. 
This reveals that several states lie between the unmitigated and ``Krylov'' values and the true ground state. 
However, the ``Krylov+'' values lie well below the first excited state, indicating that the optimized subspace state indeed manages to produce significant overlap with the true ground state. 

The phase transition at $g=0$ is characterized by order parameters which serve as interesting additional observables for which we can reuse the IC measurement samples.
While SPT phases in general are characterized by non-local order parameters, they can be expressed as local operators for the considered MPS ground states~\cite{MPS_system_first_demo}. 
$S^X =  \sum_{j=1}^N X_j / N$ signals the trivial phase. 
It is zero for states with $g < 0$ and goes to $1$ with $g \rightarrow 1$ as $4g / (1+g)^2$. Similarly, the SPT phase is characterized by  $S^{ZY} = \sum_{j=1}^N Z_j Y_{j+1} X_{j+2} Y_{j+3} Z_{j+4} / N$ which remains zero for $g > 0$ and goes to $1$ with $g \rightarrow -1$ as $-4g / (1-g)^2$.
The obtained values for $S^X$ for the energy-optimized subspace state indeed improve over the unmitigated values for high $g$ values, see second row of Fig.~\ref{fig:PSE_MPS_results}.
However, the weight-five observable $S^{ZY}$ shown in the final row of Fig.~\ref{fig:PSE_MPS_results} is more difficult to mitigate. 
While the results recover some of the theoretical behavior for $N=16$, the statistical uncertainties in the ``Krylov+'' values dominate at larger scales. 
From this, we conclude that, while the energy remains well-mitigated throughout all system sizes, the recovered subspace state does not provide sufficient overlap with the ground state to accurately reflect complex observables like $S^{ZY}$.

\textit{Discussion---}
In this work, we presented the first large-scale implementation of quantum subspace expansion powered by classical shadows. Leveraging the high sampling rates of state-of-the-art superconducting quantum processors, we collected measurements in 32,768 randomized bases per input state and subsequently classically processed over one hundred trillion individual Pauli traces. This not only represents the realization of classical shadow measurements at an unprecedented scale, but also offers a blueprint for future workflows combining quantum and high-performance classical computing.

Our results constitute a viable path to overcoming the measurement overhead of QSE, and include -- as a key technical contribution -- a complete characterization of its statistical error bars in the presence of covariances arising from the use of IC data. Furthermore, our experiments underscore the potential of QSE as a fully noise model-agnostic error mitigation technique~\cite{suchsland2021algorithmic}, which, in contrast to other methods~\cite{van2023probabilistic, kim2023evidence, fischer2024dynamical}, does not rely on an accurate device noise characterization.

Our approach could be extended by incorporating symmetry verification~\cite{bonetmonroig2018lowcosta}, powers of the root state in the spirit of virtual distillation~\cite{huggins2021virtual} and multi-reference methods~\cite{yoshioka2022generalized}. Additionally, estimators based on medians-of-means~\cite{huang2020predicting} or dual frame optimization~\cite{fischer2024dual, malmi2024enhanced, caprotti2024optimising} could further reduce the sampling costs.

It is also worth noticing that only sufficiently local observables can be estimated efficiently when single-qubit randomized measurements are employed.
For accurate ground state recovery, our technique thus requires the root state $\rho_0$ to be at most a few local operators away from the target state. 
Interestingly, this probes a regime that is complementary to that of single-shot-based subspace expansion~\cite{kanno2023quantum, robledomoreno2025}, whose success hinges on the fact that the ground state is only supported on polynomially many basis states. 
POVM measurements with less local operators, such as shallow~\cite{bertoni2024shallowshadows}, matchgate~\cite{wan2023matchgate} or global Clifford shadows~\cite{huang2020predicting} could therefore broaden the scope of our method.

\textit{Acknowledgements---}
We thank Moritz Scheer, Alberto Baiardi, and Elisa Bäumer for support on constructing MPS ground state circuits. 
We thank Guillermo García-Pérez and Oriol Vendrell for helpful discussions in the early stages of the project.
L.E.F. and D.B. acknowledge funding from the European Union’s Horizon 2020 research and innovation program under the Marie Sk\l{}odowska-Curie grant agreement No.~955479 (MOQS – Molecular Quantum Simulations). 
This research was supported by NCCR
MARVEL, a National Center of Competence in Research,
funded by the Swiss National Science Foundation (grant
number 205602) and by RESQUE, funded by the Swiss
National Science Foundation (grant number 225229).

\clearpage

\appendix

\section{Details on hardware demonstration}
\label{app_experimental_details}

\subsection{The \textit{ibm\_fez} device}
\label{app_device}

Our experimental demonstration is performed through a cloud-based access of the IBM Quantum 156-qubit device \emph{ibm\_fez} using Qiskit~\cite{qiskit2024}.
This device uses a native gate set of $\{\text{CZ},\sqrt{X},R_z\}$ on a heavy-hexagonal topology, see insets in Fig.~\ref{fig:PSE_MPS_results} for the qubit layout.
The device achieved median values of $T_1 = 136\,\upmu \text{s}$, $T_2 = 79\,\upmu \text{s}$, average gate fidelities of $f_\text{CZ} = 2.8\times 10^{-3}$ and $f_{\sqrt{SX}} = 4.4\times 10^{-4}$, and median readout errors of $1.6\%$ at the time of performing the experiments. 

\subsection{Details on circuits}
\label{app_circuits}

\begin{figure*}[t]
    \centering
    \includegraphics[width=0.88\textwidth]{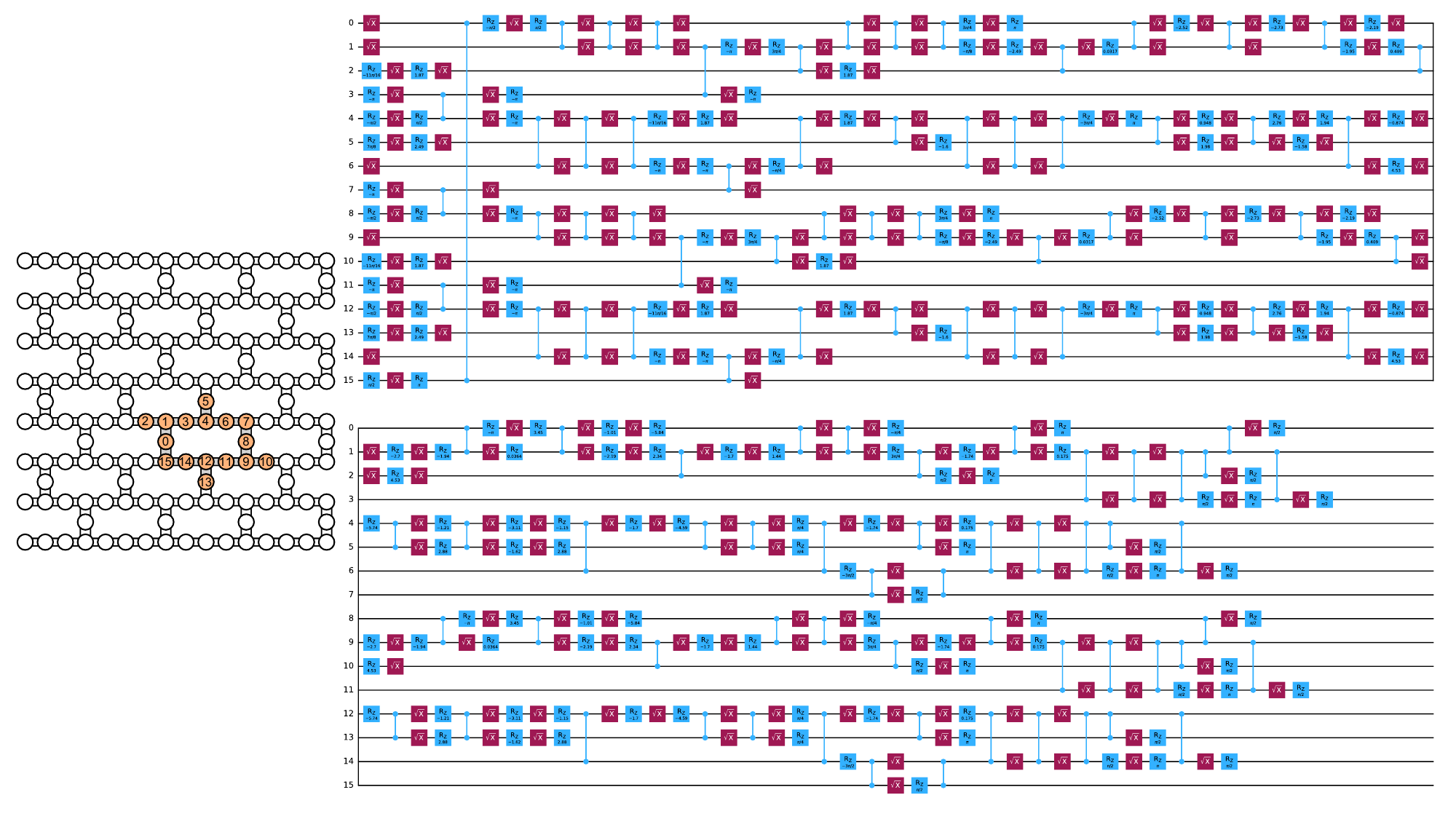}
    \caption[]{Quantum circuit that prepares an approximate ground state of the Hamiltonian from Eq.~\eqref{eq:Hamiltonian_MPS_toy_model} for $N=16$ with $g = -0.7$ through a renormalization-group-based MPS ansatz (without readout basis rotations used for classical shadow measurements). 
    The left panel shows the physical qubit layout used in the experimental demonstration. Red boxes represent single-qubit $\sqrt{X}$ gates while blue boxes represent parameterized $R_z $ gates. 
}
\label{fig:MPS_circuit_example}
\end{figure*}

For constructing the MPS ground state preparation circuits of the considered spin model we follow Ref.~\cite{scheer2025renormalization}, including the strategy for mapping circuit qubits to hardware qubits described therein, which is designed to minimize the number of required SWAP gates (and thus entangling CZ gates). 
The resulting numbers of two-qubit gates upon transpilation is summarized in Tab.~\ref{tab:details_subspace_expansion}.
As the transpiled circuits feature long idle times of several qubits, see Fig.~\ref{fig:MPS_circuit_example} for an $N=16$ circuit, we apply dynamical decoupling with an $XX$ sequence on idling qubits. 
The qubit layouts shown in Fig.~\ref{fig:PSE_MPS_results} were chosen to optimize performance based on the reported qubit quality metrics and gate fidelities.

As discussed in Ref.~\cite{scheer2025renormalization}, the circuits only approximate the true ground state of the Hamiltonian due to a fixed point approximation error in the underlying renormalization-group-based protocol. 
This approximation has been shown to be insignificant for $\abs{g}\geq 0.5$ but does introduce a noticeable error for the $\abs{g} = 0.1$ cases.
In that regard, our experiments underscore that subspace expansion not only serves the purpose of hardware error mitigation but can also overcome small algorithmic errors. 

\subsection{Classical shadow measurement}
\label{app_classical_shadows}

We perform randomized measurements known as single-qubit Pauli classical shadows~\cite{huang2020predicting}.
That is, we sample a number of $N_c$ different measurement configurations where each qubit is measured either in the $X$, $Y$, or $Z$ basis (chosen uniformly at random for each qubit), and take $N_S$ individual shots in each measurement basis. 
This has traditionally led to significant circuit overheads, as, in the most direct approach, this results in $N_c$ different quantum circuits that have to be processed by the entire hardware stack. 
Instead, we leverage a parametric circuit compilation pipeline facilitated by the \texttt{povm-toolbox} repository and by the Qiskit Runtime primitives~\cite{qiskit2024}, thus making randomized measurements significantly more efficient. 
In this approach, each basis rotation prior to readout is implemented as a parameterized sequence of $R_z(\theta_3) \times \sqrt{X} \times R_z(\theta_2) \times \sqrt{X} \times R_z({\theta_1})$, including the identity transformation for measuring in the $Z$ basis, for which the two $\sqrt{X}$ pulses cancel. 
Hence, only one pulse sequence template is processed by the device control stack and the sets of parameters representing the different readout bases only correspond to different phases on these pulses. 
This enables fast, hardware-side parameter binding that allows us to explore significantly more basis randomizations, namely $N_c = 32,768$ with $N_S = 8$, compared to previous classical shadow measurements reported in the literature with comparable qubit numbers ($N_c = 256$ with $N_S = 1024$ for 91 qubits in Ref.~\cite{fischer2024dynamical} and $N_c = 2048$ with $N_S = 1024$ for 96 qubits in Ref.~\cite{votto2025learning}.)
In total, the quantum circuit execution for all results shown in Fig.~\ref{fig:PSE_MPS_results} took $\approx 1\,\text{h}$ of wall clock time. 

A single shot of this randomized measurement procedure is distributed according to the  POVM $\{M_k\}$ with $n = 6^N$ operators that are tensor products of single-qubit operators $ M_{{k}} = M_{k_1,k_2,\dots,k_N} = M^{(1)}_{k_1} \otimes M^{(2)}_{k_2} \otimes \cdots \otimes M^{(N)}_{k_N} $ with the single-qubit POVMs $M^{(i)}_{k_i} = \{ \frac{1}{3} \ket{j}\bra{j}\}_{\ket{j}}$ where $ \ket{j} \in \{ \ket{0}, \ket{1}, \ket{+}, \ket{-}, \ket{i}, \ket{-i}\}$. 
The dual operators used for post-processing are the canonical duals for which $\{M_k\} = \{D_k\}$~\cite{fischer2024dual}, hence the dual operators retain the product structure of the POVM. 
We write all observables $O$, expansion operators $\sigma_i$, and dual operators $D_k$ as linear combinations of Pauli operators.
The traces needed to evaluate Eq.~\eqref{eqn:exp_value_subspace} then boil down to products of single-qubit traces between Pauli operators. 
This results in $\mathcal{O}(10^{12})$ single-qubit traces we process to span the ``Krylov+'' subspaces for each system size $N$ (as a sum over all $g$ values), see Tab.~\ref{tab:details_subspace_expansion}.
Note that these traces only have to be computed once and the stored results can be reused to evaluate the cost function of Eq.~\eqref{eq:PSE_constrained_optimization} for different values of $\vec{c}$ during the optimization. 

\begin{table}
\centering
\begin{tblr}{c|ccccc}
\hline \hline
     $N$ & 16 & 32 & 48 & 64 & 80 \\ \hline
     CZ gates & 108 & 216 & 324 & 432 & 540 \\
     \SetCell[r=2]{c}{\# Pauli traces for \\ PSE Krylov+ ($ \times 10^{12}$)} & \SetCell[r=2]{c} 0.4 & \SetCell[r=2]{c} 2.9 & \SetCell[r=2]{c} 11.3 & \SetCell[r=2]{c} 30.3 & \SetCell[r=2]{c} 56.3 \\
      &     &     &     &     &     \\
      ${\epsilon_\text{max} / \hat H_\text{unmit}}$ &  0.05   &  0.075   &  0.1   &  0.125   &  0.15  \\
\hline \hline
\end{tblr}
\caption{Details on the experimental demonstration. 
The number of CZ gates corresponds to the transpiled ground state preparation circuits. 
The given number of Pauli traces are the total number of single-qubit traces evaluated in post-processing for each system size. 
The allowed error $\epsilon_\text{max}$ for the constrained optimization is taken as a fixed multiple of the unmitigated energy estimate $\hat H_\text{unmit}$.
}
\label{tab:details_subspace_expansion}
\end{table}

\newpage
\section{Classical postprocessing}
\label{app_postprocessing}

\subsection{Expansion operator selection}
\label{app_expansion_ops}

Here, we detail the Pauli operators selected to construct the ``Krylov+'' subspaces of our experimental demonstration.
As mentioned in the main text, we included Pauli expansion operators $\sigma$ with weights ranging from $w=1$ to $w=5$.
For each considered weight $w$, the chosen Paulis of the subspace are listed in Tab.~\ref{tab:pauli_table}.
Note that the total dimensions of the ``Krylov+'' spaces is $L=27$ for the smallest system size $N=16$ but only $L=14$ for $N=80$. 
This choice was made to keep the post-processing feasible to run on a standard laptop, see Appendix.~\ref{app_computational_resources}. 
While this was sufficient to mitigate noise in the energy estimation of the considered spin model circuits, the dimensions of the considered subspaces could be significantly increased with high-performance computing resources, where the number of Pauli traces that need to be evaluated grows roughly quadratically in $L$. 

\small
\begin{table*}
    \fontfamily{pcr}\selectfont
    \centering
    \scalebox{0.7}
    {\begin{tblr}{|c|c|c|}
    \hline
    $N$ & $g$ & selected Paulis \\
    \hline \hline
16 & -0.9 & 
{ $Z_{4}$, $X_{7}$, $X_{3}$, $Y_{8}$, $X_{12}$, $Y_{1}X_{2}$, $Z_{5}Z_{12}$, $Z_{3}Z_{10}$, $X_{2}Z_{10}$, $Z_{7}X_{11}$, $Z_{5}X_{14}Z_{16}$, $Z_{2}X_{13}Y_{14}$, $Z_{1}Y_{2}X_{7}$, $Z_{2}Z_{4}Z_{11}$, $Y_{1}Y_{9}X_{11}$, $Y_{10}X_{13}Z_{14}Y_{15}$,\\ $Y_{2}Z_{3}X_{7}X_{11}$, $Z_{5}Y_{11}X_{12}X_{16}$, $Y_{5}Y_{9}Z_{10}Z_{16}$, $Z_{4}Z_{12}Y_{13}X_{16}$, $Y_{1}Y_{3}Z_{6}Z_{11}Y_{13}$, $Z_{1}X_{2}Y_{5}Z_{8}Y_{12}$, $Y_{1}X_{2}Y_{4}X_{7}Z_{11}$, $Y_{1}Z_{3}Z_{9}Z_{10}X_{11}$, $Z_{3}Y_{4}Y_{13}X_{14}Z_{15}$ } \\
16 & -0.5 & 
{ $X_{9}$, $Z_{9}$, $Z_{7}$, $Z_{14}$, $Y_{11}$, $Y_{6}X_{7}$, $Z_{9}Z_{16}$, $Z_{4}Z_{6}$, $Z_{9}Z_{10}$, $Z_{1}Z_{13}$, $Z_{1}Y_{5}Z_{12}$, $X_{12}Z_{13}Z_{15}$, $Y_{5}Y_{9}Y_{12}$, $Y_{10}X_{14}X_{15}$, $X_{9}X_{13}X_{14}$, $Y_{8}Y_{9}X_{11}X_{15}$, \\$Y_{1}Y_{6}Z_{7}Y_{8}$, $Y_{2}Y_{5}X_{6}Y_{8}$, $X_{2}X_{7}X_{8}Z_{9}$, $Z_{6}X_{8}X_{10}Y_{11}$, $Y_{1}Z_{2}Y_{4}Z_{14}X_{15}$, $Y_{1}Z_{4}Y_{7}Y_{13}Z_{15}$, $Z_{1}X_{4}Z_{5}X_{9}Y_{11}$, $Y_{6}Y_{7}Y_{10}Z_{11}Z_{16}$, $X_{1}Z_{4}Y_{5}Y_{7}Z_{15}$ } \\
16 & -0.1 & 
{ $Z_{7}$, $Z_{1}$, $Z_{3}$, $X_{13}$, $Z_{8}$, $Z_{1}Z_{7}$, $Z_{10}Z_{12}$, $Z_{3}Z_{4}$, $Z_{9}Z_{10}$, $Z_{13}Z_{15}$, $Z_{2}Z_{3}X_{9}$, $X_{5}X_{10}Z_{12}$, $Z_{2}X_{7}Z_{9}$, $Z_{3}Z_{5}X_{6}$, $Y_{4}Z_{5}Z_{10}$, $Z_{4}Z_{6}Z_{7}Z_{13}$, \\$X_{9}Z_{10}Z_{11}Z_{14}$, $Y_{3}Y_{5}Z_{8}Z_{15}$, $Z_{2}Y_{4}Z_{11}Z_{16}$, $X_{5}Z_{6}Z_{10}Z_{11}$, $X_{2}Y_{3}Y_{6}Y_{8}Y_{9}$, $Z_{1}Z_{3}Z_{9}Y_{13}Y_{16}$, $X_{1}Y_{4}Y_{13}Z_{14}Y_{16}$, $Y_{7}Y_{8}Z_{11}X_{13}X_{15}$, $Z_{1}Z_{5}Z_{10}Y_{14}X_{15}$ } \\
16 & 0.1 & 
{ $X_{10}$, $X_{13}$, $X_{7}$, $X_{2}$, $X_{4}$, $Z_{10}Z_{15}$, $Z_{6}Z_{8}$, $X_{4}X_{13}$, $X_{10}X_{11}$, $Z_{13}Z_{16}$, $Z_{10}Z_{11}X_{13}$, $X_{4}Z_{6}Z_{7}$, $X_{10}Z_{14}Z_{15}$, $Z_{5}X_{6}Z_{12}$, $Z_{2}Z_{10}Z_{16}$, $X_{9}Y_{10}Z_{12}X_{16}$,\\ $Z_{6}X_{9}X_{11}X_{12}$, $Z_{7}X_{9}Z_{11}Z_{12}$, $Z_{5}Z_{9}X_{15}Y_{16}$, $X_{2}X_{8}Y_{14}Y_{16}$, $Z_{2}Z_{5}Z_{12}Z_{14}X_{16}$, $Z_{1}Z_{5}X_{7}Z_{11}Z_{12}$, $Z_{2}Z_{3}X_{12}X_{13}X_{16}$, $Y_{1}Z_{5}Y_{6}X_{7}Y_{9}$, $Y_{3}Z_{6}Z_{8}X_{12}X_{16}$ } \\
16 & 0.5 & 
{ $X_{10}$, $X_{12}$, $X_{11}$, $X_{5}$, $X_{2}$, $X_{5}X_{12}$, $X_{11}X_{15}$, $X_{4}X_{12}$, $X_{10}X_{13}$, $X_{7}X_{11}$, $X_{10}X_{11}X_{15}$, $X_{4}X_{6}X_{11}$, $X_{4}X_{6}X_{7}$, $X_{6}X_{7}X_{8}$, $Z_{9}Z_{11}X_{13}$, $X_{4}X_{7}X_{13}X_{15}$,\\ $X_{2}X_{10}X_{13}X_{14}$, $X_{2}X_{8}X_{11}X_{14}$, $X_{4}X_{10}Z_{14}Z_{15}$, $X_{7}X_{8}Y_{10}Z_{14}$, $Z_{3}X_{5}Y_{10}Z_{11}X_{15}$, $X_{1}Y_{2}Y_{4}X_{12}Z_{14}$, $Z_{3}Y_{8}X_{12}Z_{14}Z_{16}$, $Z_{4}Y_{5}X_{9}Y_{12}Y_{13}$, $Z_{1}Z_{7}X_{8}Z_{14}X_{16}$ } \\
16 & 0.9 & 
{ $X_{12}$, $X_{11}$, $X_{5}$, $X_{2}$, $X_{9}$, $X_{8}X_{10}$, $X_{7}X_{10}$, $X_{6}X_{10}$, $X_{11}X_{13}$, $X_{10}X_{16}$, $X_{3}X_{8}X_{9}$, $X_{2}X_{11}X_{15}$, $X_{3}X_{7}X_{11}$, $X_{2}X_{8}X_{9}$, $X_{3}X_{7}X_{13}$, $X_{3}X_{9}X_{12}X_{13}$,\\ $X_{3}X_{9}X_{10}X_{14}$, $X_{3}X_{7}X_{8}X_{12}$, $X_{1}X_{2}X_{7}X_{16}$, $X_{6}X_{8}X_{11}X_{12}$, $X_{5}Y_{8}X_{10}X_{11}X_{16}$, $X_{2}Y_{5}X_{7}X_{13}Y_{15}$, $Y_{2}X_{8}X_{11}X_{12}X_{14}$, $Y_{2}Y_{3}Y_{6}X_{9}Y_{13}$, $X_{1}Y_{2}X_{11}Y_{12}Y_{16}$ } \\
\hline
32 & -0.9 & 
{ $Z_{11}$, $Z_{3}$, $Z_{13}$, $X_{20}$, $Z_{4}X_{7}$, $Y_{11}Z_{27}$, $Z_{19}X_{26}$, $Z_{2}Z_{23}$, $X_{20}Y_{22}Y_{24}$, $X_{0}X_{24}Z_{31}$, $Y_{7}Z_{18}Z_{27}$, $Z_{5}Z_{22}X_{31}$, $X_{0}X_{4}Z_{7}Z_{8}$,\\ $Z_{20}X_{24}Z_{27}X_{30}$, $Z_{2}Y_{4}Y_{12}X_{16}$, $Y_{2}Z_{5}X_{7}X_{31}$, $X_{14}Y_{15}X_{25}Z_{28}X_{31}$, $X_{1}Z_{6}Y_{15}Z_{16}Z_{19}$, $Z_{11}Z_{19}Z_{23}Y_{25}Y_{26}$, $Y_{7}Y_{15}X_{28}X_{30}X_{31}$ } \\
32 & -0.5 & 
{ $Z_{30}$, $Z_{12}$, $Z_{1}$, $Z_{9}$, $Z_{1}Z_{6}$, $Z_{6}Z_{9}$, $Z_{7}Z_{9}$, $X_{3}X_{22}$, $Z_{24}X_{27}Z_{29}$, $Z_{2}Z_{8}Z_{12}$, $X_{7}Z_{14}X_{30}$, $X_{9}Y_{20}X_{30}$, $Z_{5}Y_{8}Z_{12}X_{19}$,\\ $Z_{8}Y_{12}Y_{19}Z_{25}$, $Z_{1}Y_{22}Z_{26}Z_{28}$, $Y_{3}Y_{5}Y_{17}Y_{23}$, $Y_{9}Y_{14}X_{18}Y_{23}X_{24}$, $Y_{5}X_{12}Z_{16}Z_{23}Z_{26}$, $Z_{2}Y_{10}Z_{11}X_{15}X_{23}$, $X_{4}Y_{6}X_{17}Y_{19}X_{23}$ } \\
32 & -0.1 & 
{ $Z_{25}$, $Z_{27}$, $Z_{1}$, $Z_{24}$, $Z_{13}Z_{15}$, $Z_{10}Z_{11}$, $Z_{25}Z_{26}$, $Z_{16}Z_{18}$, $X_{4}Y_{12}Y_{19}$, $Z_{12}Z_{22}X_{24}$, $Z_{14}Z_{15}Y_{29}$, $Z_{1}X_{6}Z_{20}$, $Z_{9}X_{28}Y_{29}X_{30}$,\\ $Z_{19}Z_{22}Z_{28}X_{31}$, $Z_{14}Z_{20}Y_{21}Z_{24}$, $Z_{11}Y_{17}Z_{18}Z_{31}$, $Y_{8}Y_{16}Z_{26}Y_{28}X_{31}$, $Z_{1}X_{7}Y_{13}Z_{20}Z_{28}$, $Z_{2}Z_{13}Z_{24}Z_{27}Z_{31}$, $Z_{4}Y_{6}Z_{20}X_{27}Z_{28}$ } \\
32 & 0.1 & 
{ $X_{9}$, $X_{30}$, $X_{11}$, $X_{12}$, $Z_{25}Z_{26}$, $Z_{2}Z_{28}$, $Z_{9}Z_{29}$, $X_{8}X_{29}$, $Z_{3}X_{8}Z_{13}$, $X_{9}Z_{28}Z_{31}$, $X_{13}Y_{17}Y_{22}$, $Z_{1}Z_{12}Z_{28}$, $Z_{11}Z_{14}Z_{20}Z_{25}$,\\ $Y_{0}X_{11}Y_{28}Y_{31}$, $Z_{13}Z_{26}Z_{27}Z_{29}$, $Y_{14}X_{24}Y_{28}Z_{29}$, $Y_{0}Y_{1}X_{2}X_{3}Y_{16}$, $X_{1}X_{8}Y_{11}Z_{16}Z_{17}$, $X_{4}Y_{10}Y_{12}Z_{15}Y_{19}$, $Y_{6}Z_{8}Y_{19}Z_{21}Z_{24}$ } \\
32 & 0.5 & 
{ $X_{11}$, $X_{9}$, $X_{19}$, $X_{10}$, $X_{11}X_{20}$, $X_{3}X_{28}$, $X_{5}X_{19}$, $X_{4}X_{10}$, $X_{11}X_{22}X_{28}$, $X_{7}X_{17}X_{21}$, $X_{2}X_{10}X_{27}$, $X_{5}X_{6}X_{23}$, $X_{3}X_{6}X_{14}X_{16}$,\\ $Z_{2}Z_{6}Y_{22}Z_{27}$, $X_{4}X_{13}X_{17}Y_{31}$, $Y_{6}Z_{22}X_{23}Z_{30}$, $X_{1}Y_{12}Y_{20}X_{24}X_{26}$, $Z_{0}Y_{4}X_{10}Z_{13}X_{30}$, $X_{10}Z_{11}Y_{16}X_{17}X_{31}$, $Y_{4}X_{19}Z_{21}X_{22}X_{23}$ } \\
32 & 0.9 & 
{ $X_{11}$, $X_{9}$, $X_{19}$, $X_{5}$, $X_{11}X_{20}$, $X_{8}X_{19}$, $X_{1}X_{8}$, $X_{10}X_{20}$, $X_{0}X_{20}X_{30}$, $X_{1}X_{24}Y_{31}$, $Y_{3}X_{20}X_{25}$, $X_{14}Y_{18}X_{31}$, $X_{21}X_{22}X_{30}X_{31}$,\\ $Y_{3}Y_{4}Z_{18}Z_{30}$, $Y_{0}Z_{12}X_{17}Z_{21}$, $Z_{0}Z_{14}Z_{17}Y_{22}$, $X_{9}X_{10}X_{15}X_{18}X_{19}$, $X_{0}Y_{2}X_{16}X_{23}X_{31}$, $X_{0}Z_{5}X_{17}X_{18}X_{19}$, $X_{2}Y_{10}Y_{14}X_{15}X_{20}$ } \\
\hline
48 & -0.9 & 
{ $Z_{23}$, $X_{29}$, $Z_{21}$, $Y_{45}$, $Z_{38}X_{41}$, $Y_{3}Z_{28}$, $X_{16}X_{28}$, $Y_{1}Z_{2}$, $Z_{9}Y_{32}X_{34}$, $Z_{11}Z_{31}Z_{46}$, $Y_{28}Z_{32}X_{43}$, $Z_{14}Z_{28}X_{46}$, $Z_{14}Z_{18}X_{22}X_{23}$, \\$X_{15}Y_{35}X_{42}Y_{47}$, $X_{5}X_{11}Z_{13}X_{16}$, $Z_{4}Z_{16}Z_{17}X_{32}$, $Y_{5}Y_{14}Y_{15}Z_{20}X_{36}$, $Y_{8}X_{12}Z_{39}X_{45}Y_{46}$, $Y_{12}Y_{24}X_{29}Z_{36}Y_{37}$, $X_{11}X_{17}Z_{22}X_{27}Z_{30}$ } \\
48 & -0.5 & 
{ $Z_{19}$, $Z_{14}$, $Z_{10}$, $Z_{38}$, $X_{2}X_{19}$, $Y_{17}X_{23}$, $Z_{5}X_{42}$, $Z_{24}Z_{29}$, $X_{6}Z_{35}Z_{36}$, $X_{6}Y_{23}Z_{29}$, $Z_{0}Z_{4}Y_{15}$, $X_{8}Z_{10}X_{34}$, $X_{19}Y_{22}Z_{27}Z_{43}$, \\$X_{13}X_{22}Y_{24}Z_{42}^\bigstar$, $Z_{14}X_{17}Z_{20}Z_{22}$, $Z_{24}Z_{27}Y_{33}X_{46}$, $Z_{3}Z_{14}Y_{25}Z_{27}Z_{44}$, $Y_{7}Z_{11}Y_{25}X_{36}X_{38}$, $X_{2}Z_{3}X_{10}Y_{41}Y_{46}$, $X_{29}Z_{31}X_{33}Y_{41}Z_{42}$ } \\
48 & -0.1 & 
{ $Z_{33}$, $Z_{3}$, $Z_{30}$, $Z_{8}$, $Z_{40}Z_{43}$, $Z_{32}Z_{33}$, $Z_{31}Z_{35}$, $Z_{7}Z_{20}$, $Z_{0}Z_{19}Z_{45}$, $Z_{1}Y_{34}Y_{46}$, $X_{6}Y_{23}Z_{47}$, $Z_{14}X_{19}Z_{24}$, $Z_{9}Z_{16}Z_{29}Z_{31}$,\\ $Z_{2}Y_{28}Z_{35}Y_{36}$, $X_{2}Z_{13}Z_{23}Z_{47}$, $Z_{11}Z_{36}Z_{38}Y_{39}$, $Z_{5}X_{10}X_{26}Y_{33}Y_{43}$, $Z_{19}Y_{25}Y_{26}Z_{32}X_{40}$, $X_{5}Y_{7}Y_{22}Y_{28}Z_{35}$, $Z_{4}X_{5}Y_{9}Z_{41}Y_{46}$ } \\
48 & 0.1 & 
{ $X_{6}$, $X_{17}$, $X_{38}$, $X_{35}$, $Z_{8}Z_{9}$, $X_{25}Z_{37}$, $X_{4}X_{36}$, $Z_{18}Z_{42}$, $Z_{19}Z_{32}X_{45}$, $Y_{4}Z_{21}Y_{22}$, $X_{18}X_{37}X_{46}$, $X_{15}Z_{18}X_{41}$, $Y_{12}Z_{26}Z_{32}Z_{36}$,\\ $Y_{2}Y_{13}Z_{33}Z_{45}$, $X_{1}Z_{24}Y_{31}Z_{45}$, $Y_{21}Y_{22}Z_{33}Z_{40}$, $Z_{3}X_{11}X_{20}Y_{31}Y_{42}$, $Z_{5}Z_{11}Z_{18}Z_{24}Z_{30}$, $Y_{9}X_{13}X_{16}Y_{19}Z_{32}$, $X_{0}Y_{9}Z_{28}Y_{35}X_{37}$ } \\
48 & 0.5 & 
{ $X_{35}$, $X_{30}$, $X_{45}$, $X_{46}$, $X_{14}X_{19}$, $X_{9}X_{31}$, $X_{35}X_{43}$, $X_{6}X_{47}$, $X_{8}X_{29}X_{35}$, $X_{4}X_{8}X_{16}$, $X_{0}X_{13}X_{43}$, $X_{15}X_{24}X_{38}$, $X_{15}X_{19}X_{20}X_{47}$,\\ $X_{11}X_{21}X_{26}X_{46}$, $X_{15}X_{16}X_{19}X_{43}$, $X_{0}X_{12}X_{20}Y_{41}$, $X_{5}X_{20}X_{27}Z_{33}X_{36}$, $Z_{15}X_{24}X_{28}Y_{30}Z_{32}$, $X_{11}X_{13}X_{24}X_{27}Y_{29}$, $Z_{6}Z_{14}Z_{32}X_{38}Z_{47}$ } \\
48 & 0.9 & 
{ $X_{19}$, $X_{12}$, $X_{30}$, $X_{28}$, $X_{11}X_{14}$, $X_{28}X_{44}$, $X_{12}X_{19}$, $X_{5}X_{22}$, $X_{3}X_{27}X_{29}$, $X_{4}X_{7}X_{13}$, $X_{2}X_{16}X_{27}$, $X_{3}X_{7}X_{42}$, $X_{18}X_{21}X_{36}X_{45}$, \\$X_{3}X_{12}X_{25}X_{31}$, $X_{18}X_{20}X_{42}Z_{44}$, $Z_{9}Z_{22}X_{30}X_{40}$, $X_{5}Y_{12}X_{26}Z_{30}Z_{38}$, $X_{12}Z_{13}X_{19}X_{28}X_{30}$, $Z_{9}X_{14}X_{25}X_{33}Y_{44}$, $Z_{14}Y_{28}Y_{32}X_{38}Z_{47}$ } \\
\hline
64 & -0.9 & 
{ $Z_{44}$, $Z_{11}$, $Y_{27}$, $Z_{2}$, $X_{9}Z_{10}$, $X_{7}Z_{34}$, $Y_{24}Z_{26}$, $Y_{13}X_{43}$, $Y_{6}X_{13}Z_{26}$, $X_{34}Z_{45}X_{47}$, $Z_{9}X_{34}X_{43}$, $Y_{4}Z_{20}Z_{57}$, $Z_{8}X_{19}Z_{52}Z_{59}$,\\ $Z_{39}Z_{46}X_{50}Z_{60}$, $X_{0}X_{8}X_{11}X_{28}$, $Y_{43}Z_{45}Y_{59}Y_{62}$, $Z_{13}X_{15}X_{16}Z_{17}X_{46}$, $X_{18}X_{35}X_{37}X_{46}Y_{49}$, $Z_{1}X_{21}Y_{31}Y_{46}X_{47}$, $X_{28}Z_{31}Z_{32}Y_{59}X_{61}$ } \\
64 & -0.5 & 
{ $Z_{54}$, $Z_{18}$, $Z_{22}$, $Z_{39}$, $Y_{13}Y_{14}$, $Y_{12}Z_{14}$, $Y_{0}Z_{19}$, $Y_{17}Y_{41}$, $Z_{6}X_{10}Y_{21}$, $Z_{29}Y_{58}X_{61}$, $Y_{15}X_{33}Z_{62}$, $X_{9}X_{20}Z_{38}$, $Z_{5}X_{28}Z_{43}Z_{48}$,\\ $Z_{11}X_{25}Y_{53}X_{56}$, $Z_{3}X_{31}Z_{46}Z_{63}$, $Y_{4}Z_{7}Z_{42}Y_{57}$, $Y_{21}Z_{26}Y_{30}Z_{46}X_{58}$, $Z_{7}X_{49}Z_{55}X_{57}X_{59}$, $Z_{2}X_{29}X_{36}Y_{42}X_{61}$, $Y_{1}X_{7}X_{8}Z_{26}X_{39}$ } \\
64 & -0.1 & 
{ $Z_{49}$, $Z_{1}$, $Z_{36}$, $X_{18}$, $Z_{28}Z_{46}$, $Z_{12}Z_{26}$, $Z_{12}Z_{44}$, $Z_{19}Y_{53}$, $Z_{19}Z_{49}Z_{51}$, $Z_{27}X_{31}Z_{62}$, $Y_{11}Z_{54}X_{60}$, $Z_{32}Y_{46}Z_{48}$, $X_{1}Y_{24}Z_{40}X_{46}$,\\ $Z_{28}Z_{56}Y_{62}X_{63}$, $Z_{0}X_{12}Y_{19}Z_{20}$, $Y_{13}X_{21}Z_{24}X_{63}$, $X_{7}Y_{34}X_{40}X_{60}Y_{63}$, $Z_{15}Y_{43}Z_{45}Z_{61}X_{62}$, $Y_{2}X_{33}Z_{36}Z_{49}X_{57}$, $Y_{0}Z_{10}Y_{15}X_{17}Z_{49}$} \\
64 & 0.1 & 
{ $X_{62}$, $X_{39}$, $X_{46}$, $X_{14}$, $Z_{1}Z_{6}$, $Z_{57}Z_{58}$, $Z_{12}Z_{54}$, $X_{8}X_{11}$, $X_{17}X_{41}X_{55}$, $X_{6}Z_{44}X_{46}$, $X_{12}Y_{35}Y_{43}$, $Z_{2}Z_{48}Y_{62}$, $Z_{8}Z_{15}Z_{52}Z_{59}$,\\ $Y_{24}Y_{30}X_{37}Z_{38}$, $X_{21}X_{33}Y_{50}Z_{56}$, $Y_{25}X_{44}X_{46}Z_{51}$, $Y_{2}Z_{16}Z_{29}Z_{41}Z_{58}$, $Y_{22}Y_{33}X_{53}Y_{59}Z_{63}$, $X_{6}Z_{30}Z_{40}Y_{42}Z_{58}$, $Z_{5}Y_{13}Z_{19}Z_{22}Z_{43}$} \\
64 & 0.5 & 
{ $X_{19}$, $X_{36}$, $X_{38}$, $X_{18}$, $X_{30}X_{37}$, $X_{5}X_{62}$, $X_{4}X_{20}$, $X_{5}X_{56}$, $X_{21}X_{33}X_{40}$, $X_{5}X_{44}X_{58}$, $X_{34}Y_{38}X_{52}$, $Z_{9}Z_{33}Z_{41}$, $X_{13}X_{31}X_{41}X_{59}$,\\ $X_{7}X_{19}X_{53}X_{58}$, $Z_{11}Z_{18}Z_{52}Z_{61}$, $Y_{7}Z_{17}Z_{38}X_{45}$, $X_{27}Z_{37}X_{46}Y_{49}X_{50}$, $X_{5}X_{15}Y_{16}Y_{19}X_{34}$, $Z_{1}X_{20}X_{25}Z_{33}Y_{63}$, $X_{16}X_{20}Y_{23}X_{39}Y_{62}$} \\
64 & 0.9 & 
{ $X_{16}$, $X_{18}$, $X_{36}$, $X_{13}$, $X_{3}X_{16}$, $X_{14}X_{25}$, $X_{53}X_{57}$, $X_{30}X_{59}$, $X_{5}X_{14}X_{57}$, $X_{18}X_{28}X_{53}$, $X_{20}X_{33}X_{58}$, $X_{5}X_{21}X_{40}$, $X_{0}X_{3}X_{44}X_{48}$,\\ $X_{9}X_{46}X_{47}Z_{58}$, $X_{4}Z_{29}X_{50}X_{54}$, $Y_{15}Y_{33}Z_{35}Z_{59}$, $Y_{2}Z_{26}X_{48}Z_{54}X_{62}$, $Z_{0}X_{5}Z_{8}Z_{10}X_{28}$, $Y_{1}Y_{3}X_{4}Y_{32}Z_{62}$, $Y_{34}Z_{36}Y_{50}Z_{62}X_{63}$} \\
\hline
80 & -0.9 & 
{ $Z_{19}$, $Y_{7}$, $Y_{4}$, $Z_{0}X_{40}$, $Z_{6}Z_{48}$, $X_{39}Z_{46}X_{54}$, $X_{16}Z_{50}Z_{59}$, $Y_{5}Z_{30}X_{41}Y_{52}$, $X_{36}X_{48}X_{59}Y_{65}$, $X_{13}X_{38}Z_{44}Z_{64}X_{68}$, $Z_{5}Y_{20}X_{21}Z_{63}Z_{72}$, $X_{2}Z_{9}Y_{22}Y_{56}Z_{70}$} \\
80 & -0.5 & 
{ $X_{5}$, $Z_{4}$, $Z_{6}$, $Z_{11}Y_{24}$, $Z_{4}Z_{72}$, $X_{40}Z_{44}Z_{53}$, $Z_{2}Z_{8}Z_{15}$, $Y_{19}X_{50}X_{64}Z_{71}$, $X_{5}Z_{14}X_{31}X_{57}$, $X_{17}X_{21}X_{50}Z_{62}Y_{65}$, $Z_{8}X_{14}X_{18}X_{38}Z_{47}$, $Y_{6}Z_{26}X_{29}Z_{46}X_{73}$} \\
80 & -0.1 & 
{ $Z_{1}$, $Z_{24}$, $Z_{4}$, $Z_{2}Z_{11}$, $Z_{40}Z_{46}$, $Z_{33}Z_{52}Z_{59}$, $Z_{35}X_{49}X_{75}$, $X_{3}X_{10}X_{16}X_{79}$, $X_{36}Z_{45}Y_{55}Y_{79}$, $Z_{0}Z_{22}Y_{51}Y_{65}Z_{68}$, $Y_{8}Z_{26}X_{31}X_{52}Z_{60}$, $Y_{3}Z_{33}X_{56}Z_{71}X_{79}$} \\
80 & 0.1 & 
{ $X_{25}$, $Z_{4}$, $Z_{17}$, $Z_{53}Z_{60}$, $X_{55}X_{71}$, $Z_{27}Z_{33}Y_{40}$, $X_{1}Y_{13}Y_{20}$, $X_{19}X_{30}Y_{36}X_{63}$, $X_{24}Z_{60}Z_{61}Y_{66}$, $Y_{15}Y_{64}Z_{65}Z_{73}Z_{76}$, $Z_{7}Y_{16}X_{54}Z_{60}X_{65}$, $Y_{6}Y_{11}Y_{26}Z_{45}Z_{62}$} \\
80 & 0.5 & 
{ $X_{21}$, $X_{5}$, $X_{4}$, $X_{13}X_{27}$, $X_{21}X_{55}$, $X_{8}X_{19}X_{60}$, $X_{36}X_{51}X_{78}$, $Z_{5}X_{14}Z_{45}X_{51}$, $X_{5}X_{10}Z_{22}Z_{77}$, $Z_{6}X_{12}Y_{29}Z_{39}X_{56}$, $Z_{6}X_{8}X_{24}X_{37}X_{56}$, $Z_{29}Y_{43}X_{61}Z_{71}Y_{77}$} \\
80 & 0.9 & 
{ $X_{22}$, $X_{20}$, $X_{17}$, $X_{65}X_{79}$, $X_{27}X_{67}$, $X_{17}X_{39}X_{77}$, $X_{18}X_{57}X_{77}$, $X_{0}X_{20}X_{41}X_{67}$, $Z_{41}X_{44}X_{54}X_{78}$, $Z_{41}Y_{46}X_{56}X_{66}X_{68}$, $Y_{5}X_{16}X_{38}Z_{54}X_{73}$, $Y_{6}Z_{18}X_{37}X_{53}Y_{63}$} \\
\hline
\end{tblr}}
\caption{
Selected Pauli expansion operators of the ``Krylov+'' expansion set for each values of system size $N$ and Hamiltonian parameter $g$ considered in our experimental demonstration. 
}
    \label{tab:pauli_table}
\end{table*}

\subsection{Computational resources}
\label{app_computational_resources}
Our classical post-processing is run on a standard laptop with an Intel Core Ultra 9 185H CPU (16 cores, 22 threads, up to 5.1 GHZ boost frequency) leveraging the Numba Python package~\cite{lam2015numba} for high-performance, parallelized Python code. 
The computational bottleneck of our method is the evaluation of the Pauli traces between subspace expansion operators and measurement dual operators summarized in Tab~\ref{tab:details_subspace_expansion}. 
In total, the post-processing for our experimental demonstration took $\approx 20\,\text{h}$. 

\subsection{Constrained optimization}
\label{app_constrained_optimization}

\begin{figure}
    \centering
    \includegraphics[width=\linewidth]{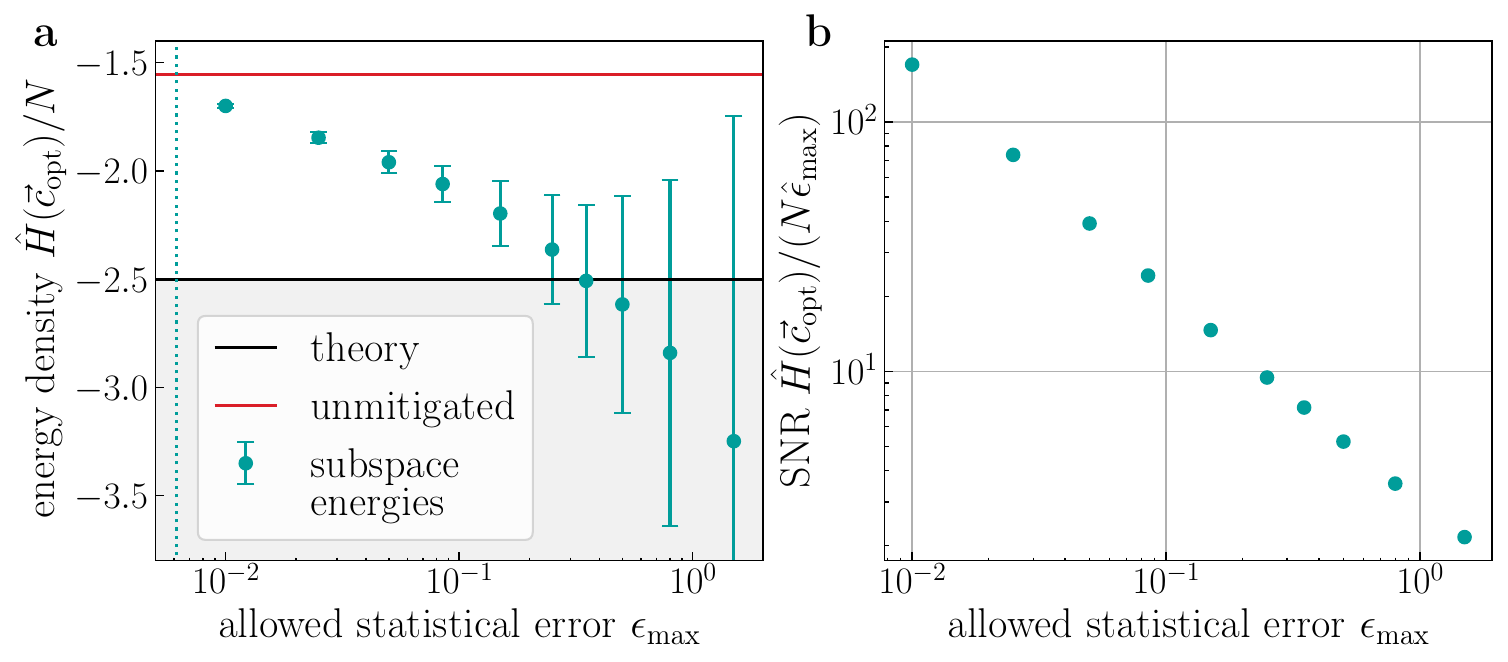}
    \caption[]{Bias-variance-tradeoff in the constrained optimization problem. 
    \textbf{a)} Obtained subspace energy density when solving the constrained optimization for increasing error budgets $\epsilon_\text{max}$. 
    The horizontal lines are the unmitigated energy estimate $\hat H_\text{unmit}$ (red) and the true ground state energy (black), while the vertical dashed line signals the statistical error of $\hat H_\text{unmit}$.
    \textbf{b)} Decrease of the signal-to-noise-ratio (SNR) of the mitigated energy density with $\epsilon_\text{max}$.
    Data corresponds to experiments with $N=48$ and $g=-0.5$.
    }
\label{fig:bias_variance_tradeoff}
\end{figure}

Here, we present details on how the constrained optimization presented in Eq.~\eqref{eq:PSE_constrained_optimization} is carried out in our experimental demonstration. 
The central hyperparameter in our procedure is the allowed maximal statistical error $\epsilon_\text{max}$, which controls a bias-variance-tradeoff. 
To illustrate this, we show an example of how the achieved subspace energy gets reduced as $\epsilon_\text{max}$ is increased, see Fig.~\ref{fig:bias_variance_tradeoff}\textbf{a}.
When the allowed error becomes too large, eventually the subspace energy can violate the variational principle, while the physical region remains within the error bars. 
This manifests in the signal-to-noise ratio of the energy density ${\hat H(\vec c_\text{opt}) / (N \hat \epsilon_\text{max}})$ which decreases roughly as a power law with $\epsilon_\text{max}$, see Fig.~\ref{fig:bias_variance_tradeoff}\textbf{b}.
This motivates choosing the allowed error in our experimental demonstration as a fixed multiple of the unmitigated signal of the energy estimation $\hat H_\text{unmit} = \hat H \left( \left(1, 0, \dots\right)\right)$. 
This ratio ranges from $5\,\%$ for $N=16$ to $15\,\%$ for $N=80$ to account for the increase in noise with the system size, see Tab.~\ref{tab:details_subspace_expansion}. 
As an initial point of the COBYLA optimization, we choose the parameters $\vec c = \left( 1, 0, 0,\dots\right)$ that correspond to the unmitigated base state $\rho_0$.
We find that the optimization converges with, on average, 1438 iterations. 

\begin{figure}
    \centering
    \includegraphics[width=\linewidth]{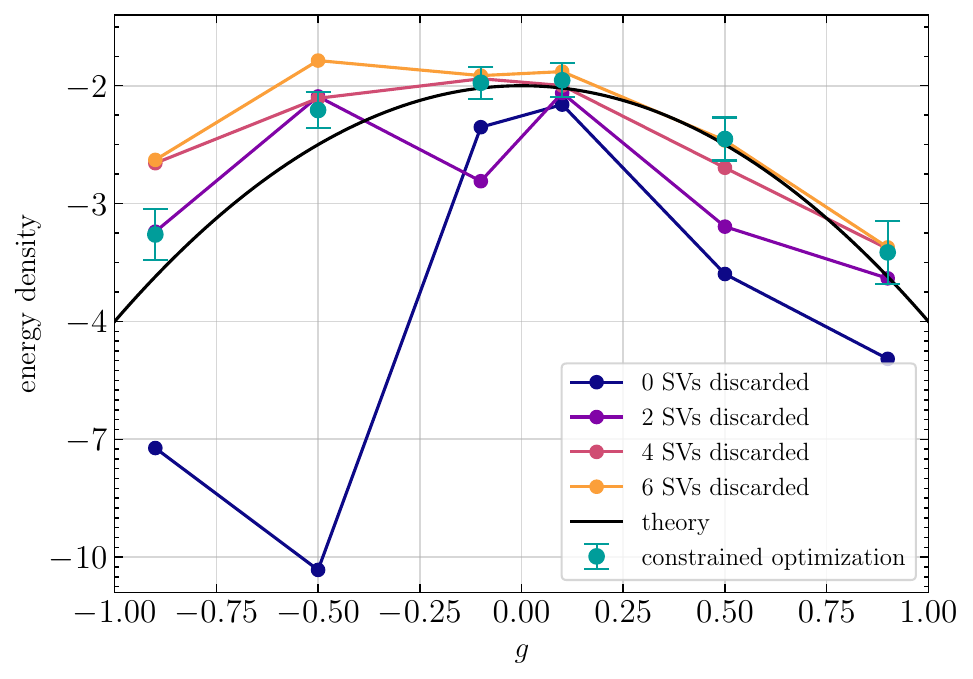}
    \caption[]{Suppressing non-physical energies through regularization of the overlap matrix on experimental data for $N=48$. 
    Data points with solid lines show the smallest pseudoeigenvalue when discarding the indicated number of the smallest singular values (SVs). 
    Data points with error bars correspond to the ``Krylov+'' subspace results from Fig.~\ref{fig:PSE_MPS_results}.
    Note the non-linear y-scale for better visibility.}
\label{fig:regularization_comparison}
\end{figure}

Besides the constrained optimization, another way to deal with statistical uncertainties in quantum subspace expansion are regularization techniques as suggested in, e.g., Ref.~\cite{urbanek2020chemistry}. 
This approach aims to overcome potential ill-conditioning of the overlap matrix $\mathcal{S}$ and solves the generalized eigenvalue problem by computing the lowest eigenvalue of $\tilde{\mathcal{S}}^{-1} \mathcal{H}$ with a regularized overlap matrix $\tilde{\mathcal{S}}$. 
$\tilde{\mathcal{S}}^{-1}$ may be computed based on a singular value decomposition $\mathcal{S} = U \mathfrak{S} V$, where $\mathfrak{S}$ is a diagonal matrix with singular values $s_i$. 
The problem is regularized by setting the diagonal entries in $\mathfrak{S}^{-1} = \text{diag}(1/s_1, \dots 1/s_L)$ to zero for the smallest singular values and computing $\tilde{\mathcal{S}}^{-1} = V^T \mathfrak{S}^{-1} U^T$.
This enables the suppression of non-physical energies in the subspace states as exemplified on the experimental data of the ``Krylov+'' subspace for $N=48$, see Fig.~\ref{fig:regularization_comparison}. 
We find that up to six singular values need to be discarded (out of a 22-dimensional subspace) to respect the variational principle, underscoring the ill-conditioning of the generalized eigenvalue problem.
Moreover, in contrast to our constrained optimization procedure, this method does not provide tunable statistical error bars. 

\end{document}